\documentclass[a4paper,11pt]{article}
\usepackage{pos}
\usepackage{graphicx}
\usepackage{caption}
\usepackage{subcaption}
\usepackage{siunitx}
\usepackage{wrapfig}

\newlength{\bibitemsep}
\newlength{\bibparskip}
\let\oldthebibliography\thebibliography
\renewcommand\thebibliography[1]{%
  \oldthebibliography{#1}%
  \setlength{\parskip}{\bibitemsep}%
  \setlength{\itemsep}{\bibparskip}%
}

\title{Development of a scintillation and radio hybrid detector array at the South Pole}
 \ShortTitle{Development of a scintillation and radio hybrid detector array at the South Pole}
 
\author{The IceCube Collaboration \\{\normalsize \normalfont(a complete list of authors can be found at the end of the proceedings)}}

\emailAdd{marie.oehler@kit.edu}
\emailAdd{roxanne.turcotte@kit.edu}

\abstract{At the IceCube Neutrino Observatory, a Surface Array Enhancement is planned, consisting of 32 hybrid stations, placed within the current IceTop footprint. This surface enhancement will considerably increase the detection sensitivity to cosmic rays in the \SI{100}{\tera\eV} to \SI{1}{\exa\eV} primary energy range, measure the effects of snow accumulation on the existing IceTop tanks and serve as R\&D for the possible future large-scale surface array of IceCube-Gen2. Each station has one central hybrid DAQ, which reads out 8 scintillation detectors and 3 radio antennas. The radio antenna SKALA-2 is used in this array due to its low-noise, high amplification and sensitivity in the 70-350 MHz frequency band. Every scintillation detector has an active area of \SI{1.5}{\meter\squared} organic plastic scintillators connected by wavelength-shifting fibers, which are connected to a silicon photomultiplier. The signals from the scintillation detectors are integrated and digitized by a local custom electronics board and transferred to the central DAQ. When triggered by the scintillation detectors, the filtered and amplified analog waveforms from the radio antennas are read out and digitized by the central DAQ. A full prototype station has been developed and built and was installed at the South Pole in January 2020.
It is planned to install the full array by 2026.
In this contribution the hardware design of the array as well as the installation plans will be presented.

\vspace{4mm}
{\bfseries Corresponding authors:}
Marie Oehler$^{1*}$, Roxanne Turcotte-Tardif$^{1}$\\
{$^{1}$ \itshape Karlsruhe Institute of Technology, D-76021 Karlsruhe, Germany}\\[4mm]
$^*$ Presenter

\FullConference{37$^{\rm{th}}$ International Cosmic Ray Conference (ICRC 2021)\\
		July 12th -- 23rd, 2021\\
		Online -- Berlin, Germany}

}

\begin{document}
\maketitle

\section{Introduction}

The IceCube Neutrino Observatory, an in-ice Cherenkov light detector located deep in the ice at the geographic South Pole, is designed to detect astrophysical neutrinos ~\cite{Aartsen_2017}.
Its surface component, IceTop, consists of 81 pairs of ice Cherenkov tanks, acting as a veto for muons entering the deep detector and as a \SI{1}{\square\kilo\meter} cosmic-ray detector.
Over the years, snow accumulated on the tanks ($\approx \SI{20}{\centi\meter\per year}$), leading to a decrease in resolution as well as an increase in threshold~\cite{snowOnIT}.
In order to investigate the impact of snow on the tanks and to improve the cosmic-ray measurements, an array of scintillation panels will be deployed over the next years, covering the whole IceTop footprint~\cite{scinticrc2017, layoutIcrc2019}. 
Radio antennas are added to this array in order to improve measurements of $X_\text{max}$ as well as improve the absolute energy scale, and thus refining measurements of the cosmic-ray composition.
These detectors will be grouped into 32 stations, each station consisting of 8 scintillation panels and 3 radio antennas, read out by a central DAQ.

Two R\&D scintillator stations with different designs were deployed in January 2018 and performed well~\cite{scinticrc2017, scinticrc2019}.
One of these stations was upgraded with two radio antennas in January 2019~\cite{radioIcrc2019}.
Using these experiences, a new prototype station combining and improving the previous iterations was designed.
This prototype station was deployed at the South Pole in January 2020, replacing the old stations.
With it, the first coincident measurements of cosmic-ray air-showers using scintillation detectors, radio antennas and IceTop were obtained~\cite{Hrvoje}.
This work will present the hardware design (sec.~\ref{sec:hardware}) and the performance (sec.~\ref{sec:performance}) of this prototype station as well as the hardware improvements for the full deployment (sec.~\ref{sec:improvements}).

\section{The hardware} \label{sec:hardware}

\subsection{The data acquisition system - TAXI v3.0}
\begin{figure}[!htb]
\centering
\includegraphics[width=\linewidth]{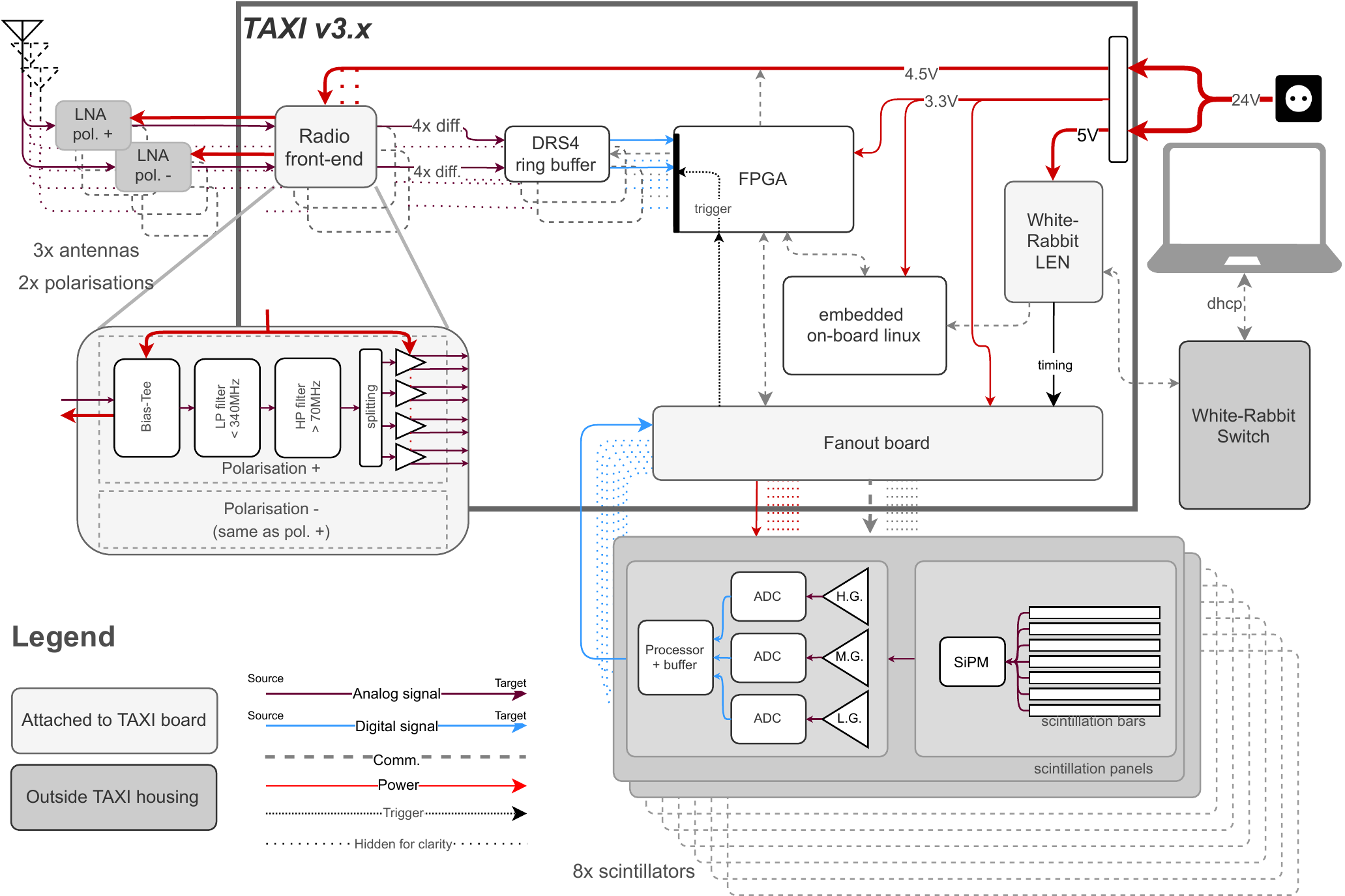}
\caption{Block diagram of all electronics for one station. 
This includes the TAXI board, the WR system, the radioTads, the fanout board, the MicroDAQs and the LNAs.
}
\label{fig:blockdiagram}
\end{figure}

The data acquisition system is an adapted version of the Transportable Array for eXtremely large area Instrumentation studies (TAXI) ~\cite{TAXI}, see fig.~\ref{fig:blockdiagram}. 
The current hardware deployed at the South Pole is TAXI v3.0.
The TAXI board has an embedded Linux operating system running on a microprocessor (\textit{Stamp9G45} by TasKit GmbH) and a FPGA (\textit{SPARTAN-6} by XILINX) for processing the signals.
The White Rabbit (WR) system provides nanosecond timing~\cite{wroriginal}.
The TAXI is powered with a double input of \SI{24}{\volt} nominal,
and the power is converted to lower voltages on the board, as required.
TAXI also provides power, communication and timing for the scintillation detectors, which are described in sec. \ref{sec:scintHardware}.

The analog radio signal output from the radioTad (sec.~\ref{sec:radiotad}) is fed into one of the three DRS4 sampling chips ~\cite{drs4} on the TAXI board and is then converted into a digital signal by a 8-channel, 14 bit, analog-to-digital converter (ADC) \textit{LTM9007IY-14}.
The DRS4 sampling chip is a ring buffer chip composed of 8+1 channels with 1024 sampling cells each. The signal of first polarisation is fed into the first four channels of the chip and the second polarisation in the next four channels.
This allows to set trace lengths of \SI{1024}{\nano\second}, \SI{2048}{\nano\second} or \SI{4096}{\nano\second} with a sampling rate of \SI{1}{\giga\hertz}.

The radio and scintillator data is sent via a \SI{1}{Gb} fiber link to the IceCube Lab (ICL) using a WR layer.
Before shipment, the TAXI board, the WR node (WR-LEN), the fanout board and the radio front-end electronics are placed in a metal housing with connectors for the detectors as well as power, communications and timing to the IceCube Lab.

\subsection{Scintillator hardware}
\label{sec:scintHardware}

The different designs of the scintillation detectors (scintillators) of the previous stations have been merged.
Each detector weighs less than \SI{50}{\kilogram} and has an active area of \SI{1.5}{\meter\squared}.
It consists of 16 plastic scintillation bars with a width of \SI{5}{\centi\meter} and a reflective coating of TiO$_2$ produced by FNAS-NICADD.
Wavelength-shifting fibers (\textit{Y-11(300)} by Kuraray) are routed through two holes in the scintillation bars to a Silicon Photomultiplier (SiPM) (\textit{13360-6025PE} by Hamamatsu).
Since SiPMs have a temperature dependent behavior, a temperature sensor, placed close to the SiPM, monitors the temperature during operation.
The SiPM signals are fed to a microprocessor-based board called Scintillator MicroDAQ v4.1 (MicroDAQ).
To increase the dynamic range, each MicroDAQ has three amplification channels.
The factor between high and medium amplification is $\approx 6.5$ and the factor between high and low amplification is $\approx 77$.
The charge of the scintillator signal is evaluated from the shaped pulse, read out by 12 bit ADCs.
Due to the WR system, the start time of the pulse can be determined with nanosecond accuracy.
The digitized data is subsequently transferred to TAXI.
When the scintillators have a signal over a pre-defined threshold value, a trigger signal is sent to TAXI to trigger the readout of the radio antennas.
The scintillators are connected to the TAXI DAQ via the scintillator fanout board. 
This board switches the power of the scintillators, enables the reprogramming of the MicroDAQs and forwards the communication and triggers.

\subsection{Radio hardware} \label{sec:radioHardware}
\subsubsection{SKALA-2 antennas}
Dual polarisation Log-Periodic Dipole Array (LPDA) antennas developed by the SKA collaboration are used as detectors for the radio emission from air showers~\cite{SKA}.
Each polarisation of the antenna has a dedicated Low Noise Amplifier (LNA). These antennas can function from 50\,MHz to 650\,MHz, however, for the surface array, the frequencies above 350\,MHz are filtered (sec.~\ref{sec:radiotad}) due to the 360.2\,MHz frequency used for air traffic at the deployment site ~\cite{commSP}.
The LNA, located directly at the top of the antenna, pre-amplifies the signal with a gain of \SI {40}{\decibel}. Its noise figure is as low as \SI{40}{\kelvin}, which lies below the expected galactic and thermal noise ~\cite{SKALAV2_LNA}. 
This was confirmed by an in-situ measurement of the galactic background noise as shown in section~\ref{sec:resultsProto}. 

To avoid snow coverage of the antennas, they are placed on a mounting structure.
Due to the snow accumulation on site, the structure needs to be raisable.
To accomplish this at small cost while avoiding reflection off a metallic structure, sapele mahagony (\textit{Entandrophragma cylindricum}) wood is used for the structure.
The baseplates are fixed to the ground with four \SI{1}{\meter} long snow spikes.
Four \SI{1}{\meter} dowels connect the baseplates with the square shaped "brace" of the structure, which allows the raising of the antenna by attaching extension dowels.
All parts of the mount were tested in the laboratory at temperatures down to \SI{-70}{\degreeCelsius} to ensure its resistance in the harsh South Pole environment. The antenna itself was previously tested at South Pole~\cite{radioIcrc2019}.

\begin{figure}[htb]
\centering
\begin{subfigure}{.35\textwidth}
  \centering
  \includegraphics[width=0.8\linewidth]{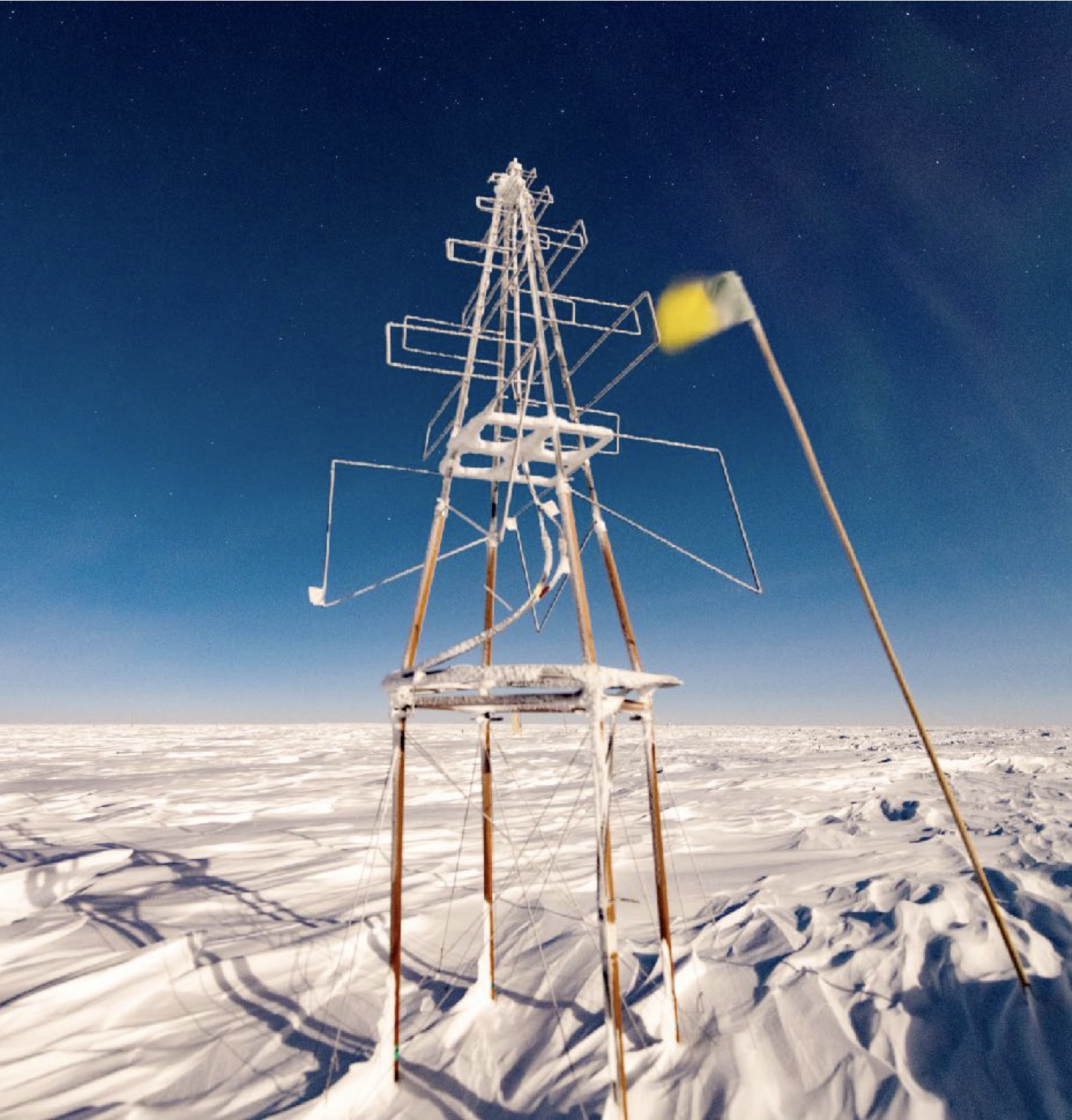}
  \caption{}
    \label{fig:antenna}
\end{subfigure}%
\begin{subfigure}{.35\textwidth}
  \centering
  \includegraphics[width=0.8\linewidth]{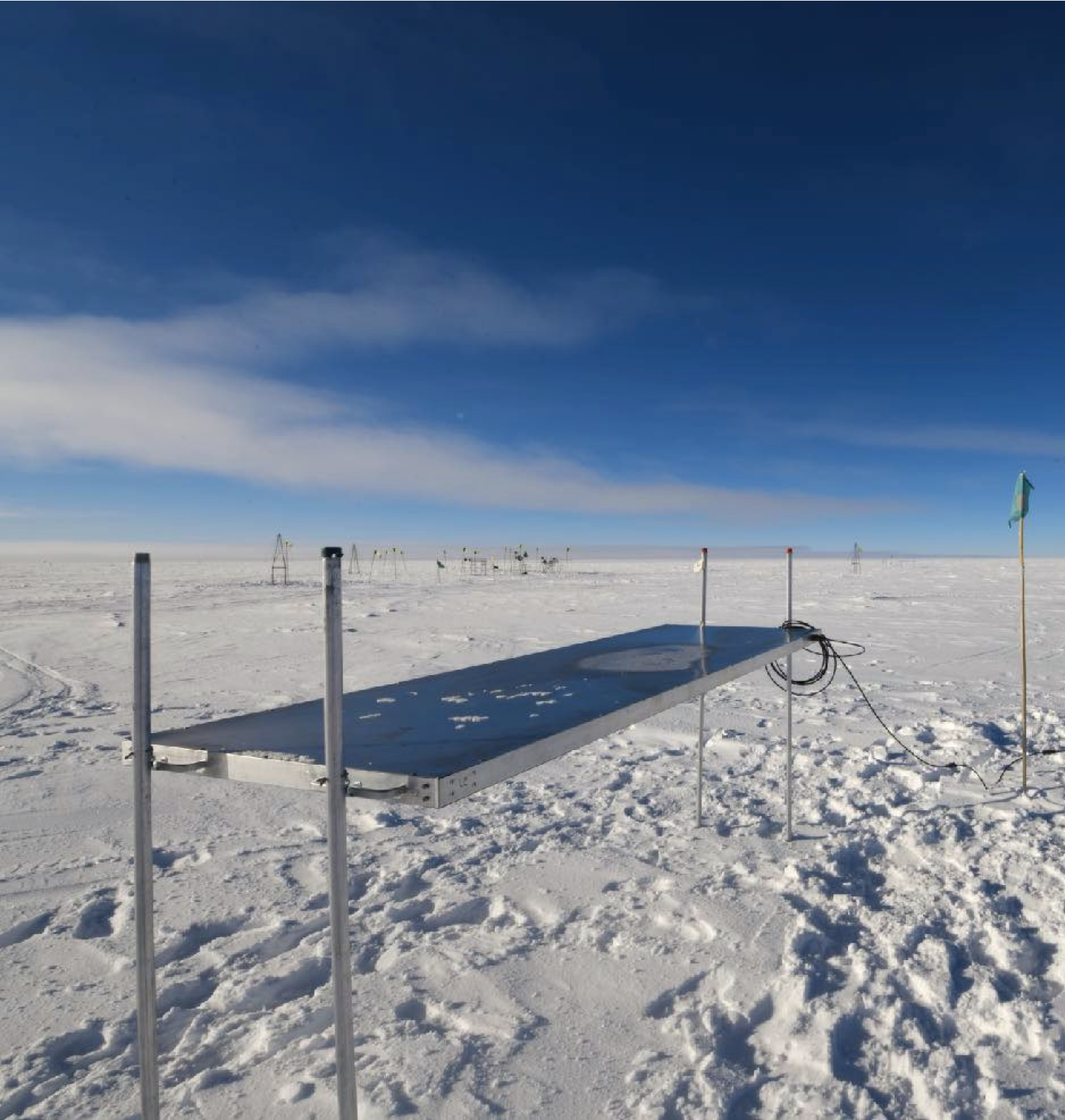}
  \caption{}
  \label{fig:scint}
\end{subfigure}
\caption{In a) an elevated SKALA-2 antenna and in b) an elevated scintillation panel is shown. Both are part of the deployed detectors in January 2020 and are both elevated above the ground and raisable to avoid snow coverage. Picture credit: Y. Makino, IceCube/NSF.}
\label{fig:detectors}
\end{figure}

\subsubsection{RadioTad - Radio front-end electronics} \label{sec:radiotad}
The radioTad processes the single-ended analog signal input from one LNA to a filtered differential analog output signal. The schematic of this board is shown on the left side of fig.~\ref{fig:blockdiagram}. There is one board for each antenna, therefore the board has two identical sections, one for each polarisation. 
For sake of clarity, only one polarisation is depicted.
A low-pass filter \textit{ULP-340+} and a high-pass filter \textit{SXHP-48+} set the bandwidth to a nominal 70-350\,MHz. A bias-tee \textit{JEBT-4R2G+} transfers the power to the LNA while receiving the signal in the same cable. The signal is then fanned-out and fed into four parallel amplifiers \textit{THS4508}, which amplify the signal by 2-3\,dB and convert it to differential pairs. 
The connections between the radioTads and the TAXI electronic box are made via 10\,cm \textit{LMR-240} cables with SMA connectors, 50\,m \textit{LMR-400} cables then connect the TAXI electronic box to a 1.5\,m \textit{LMR-240} cable attached to each LNA.
This design is based on the first version~\cite{radioIcrc2019}, with the DC-DC converter now directly incorporated in the TAXI system.

\section{Commissioning at the South Pole}\label{sec:performance}
\subsection{The deployment}

In January 2020, a first complete prototype station was deployed with 3 antennas, 8 scintillation panels and a TAXI v3.0 DAQ located in an elevated electronic box.
Fig.~\ref{fig:antenna} shows one of the deployed antennas and its mounting structure standing in the Antarctic winter and fig~\ref{fig:scint} shows one of the elevated scintillator panels.
The layout of the deployed station is based on the station layout presented in \cite{layoutIcrc2019}.

\subsection{Measurements from the deployed detectors}
\label{sec:resultsProto}
\subsubsection{Scintillation detectors}

\begin{figure}[htb]
\centering
\begin{subfigure}{.5\textwidth}
  \centering
  \includegraphics[width=0.9\linewidth]{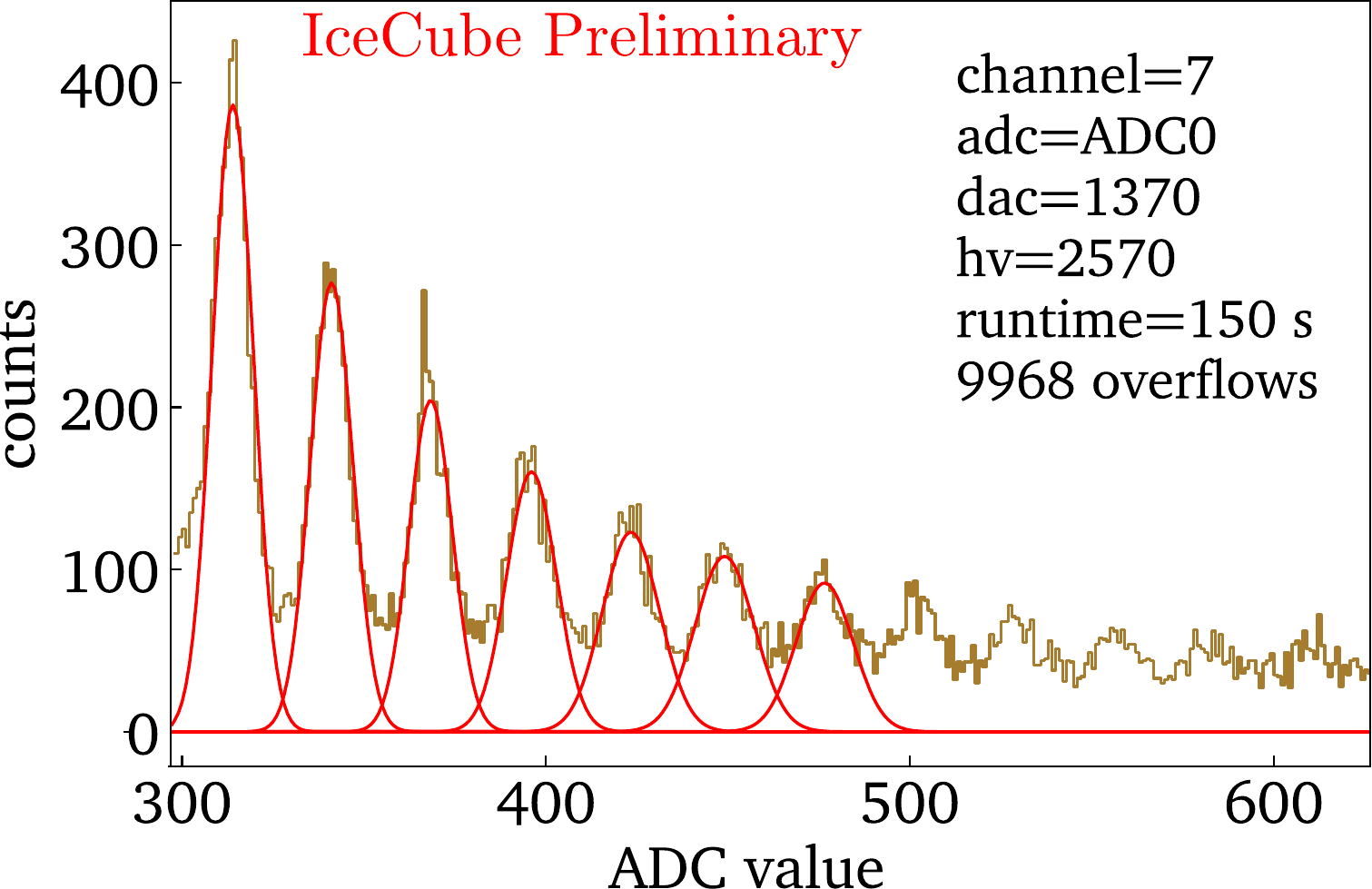}
  \caption{}
    \label{fig:sipmgain}
\end{subfigure}%
\begin{subfigure}{.45\textwidth}
  \centering
  \includegraphics[width=0.80\linewidth]{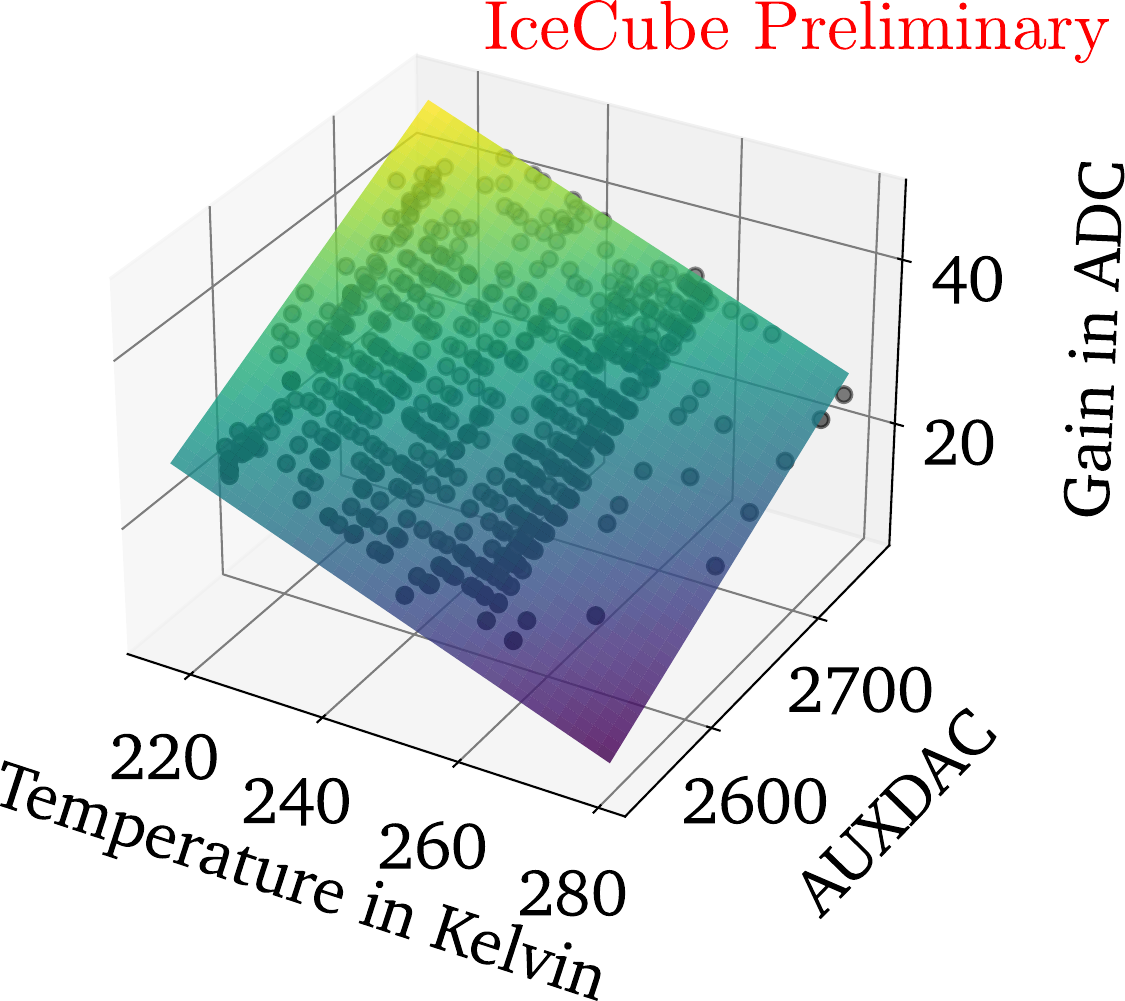}
  \caption{}
  \label{fig:gainhvtemp}
\end{subfigure}
\caption{a) The SiPM gain is evaluated from fits to single P.E. peaks. Drawn in gold is the measured data and in red, the Gaussian fits. b) The SiPM gain as function of temperature and voltage (AUXDAC).}
\label{fig:gain}
\end{figure}

The gain of the SiPM varies with temperature and applied voltage.
It can be evaluated from the distance between the single photo-electron (P.E.) peaks in the charge histogram.
\begin{wrapfigure}[11]{l}{0.42\textwidth}
\vspace{-10pt}
\centering
\includegraphics[width=0.42\textwidth]{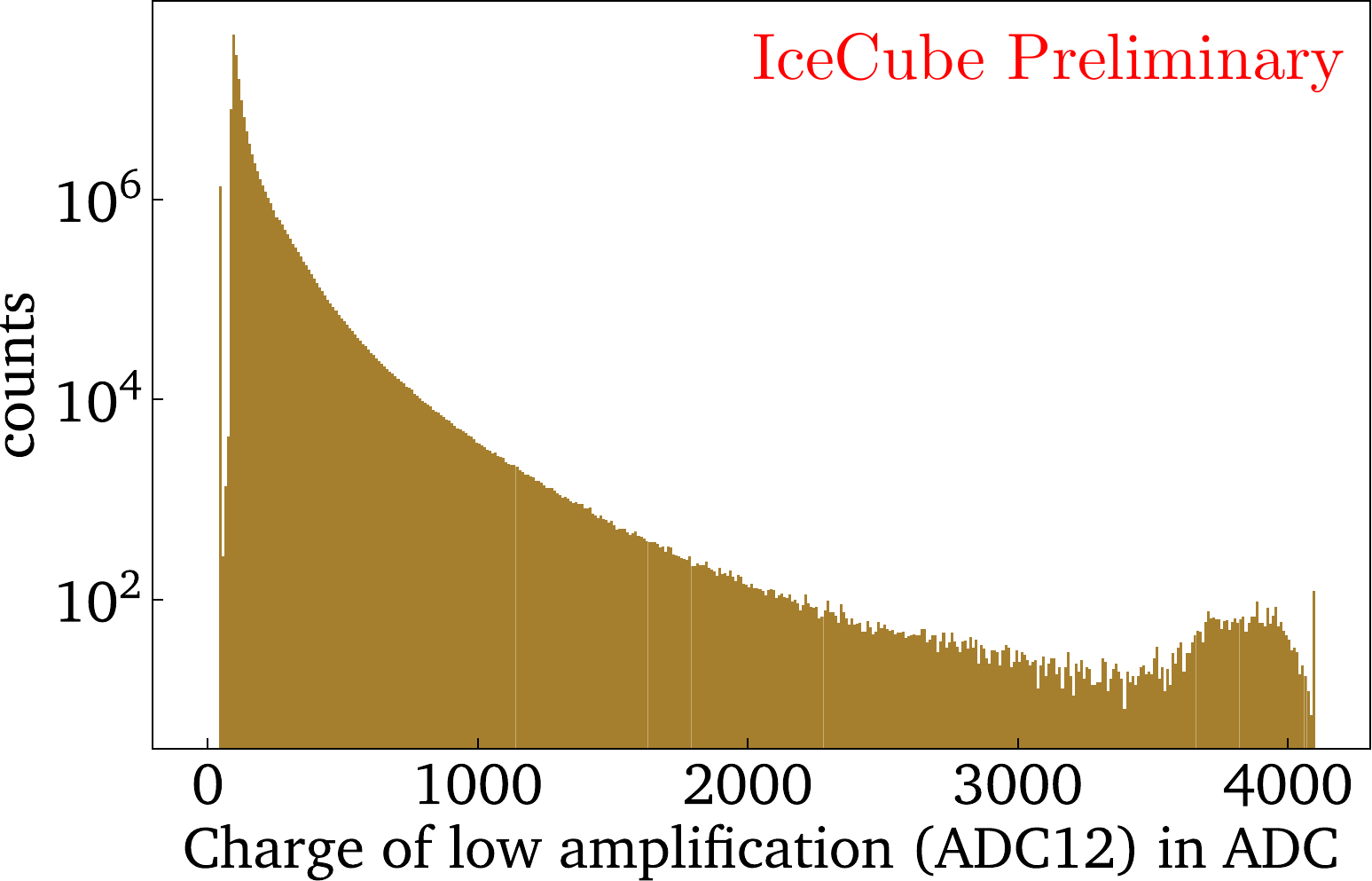}
\caption{Cumulative charge histogram of the low amplification channel. After around 3500 ADC units, a saturation is visible.}
\label{fig:saturation}
\end{wrapfigure}
As shown in fig.~\ref{fig:sipmgain}, the positions of the peaks are determined by Gaussian fits.
By repeating this measurement at different temperatures and applied voltages, the gain as a function of temperature and voltage (AUXDAC) was determined, see figure \ref{fig:gainhvtemp}.
This relation can be used to stabilize the gain during operation, so the SiPM voltage is adjusted to the temperature to keep the gain constant.

In fig.~\ref{fig:saturation} the charges measured with the low amplification channel between December 31, 2020 and April 1, 2021 are plotted as a histogram.
The pile-up visible at the end of the histogram indicates a saturation of the electronics for very high-energy air-showers.
To increase the dynamic range of the detectors to this energy range, the amplification of the low amplification channel can be reduced for the full deployment.

\begin{wrapfigure}[13]{r}[0pt]{0.60\textwidth}
\vspace{-35pt}
\centering
\includegraphics[width=0.60\textwidth]{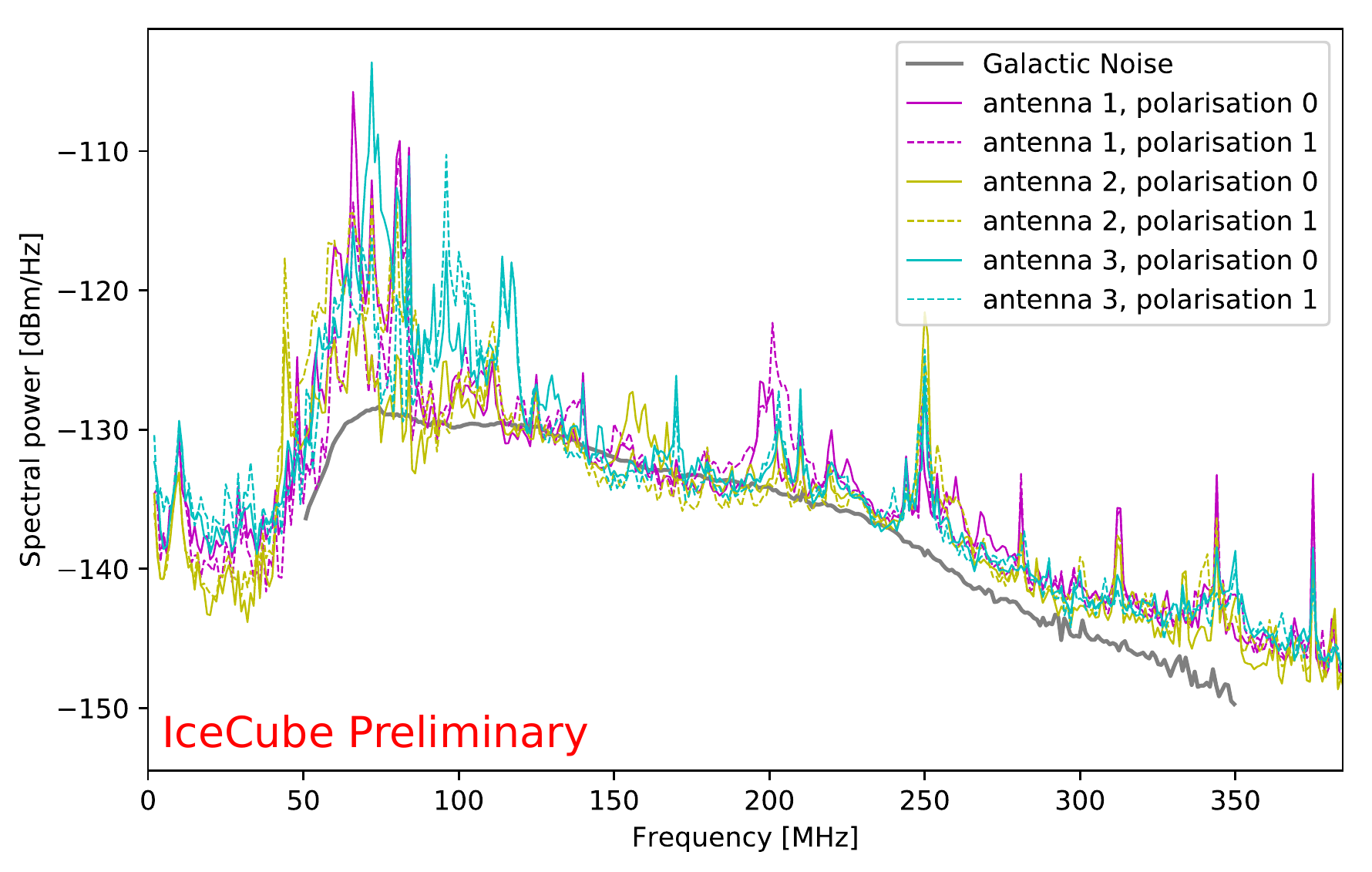}
\caption{One day median spectrum of the background (17th April 2020) of the prototype station for each antenna and polarisation. Drawn in black is the expected galactic noise from the Cane model convoluted with the electronic response~\cite{Alan}.}
\label{fig:spectrum}
\end{wrapfigure}%
\subsubsection{Radio detectors}
The median spectrum of one day of radio background data is shown in fig.~\ref{fig:spectrum}. All three antennas and their polarisations are depicted here. One can notice that the spectrum is relatively stable between the antennas. The black line denotes the galactic noise based on the Cane model and convoluted with the electronic chain~\cite{Alan}. It can be seen that, as desired, the galactic noise corresponds to the noise floor of the radio system.\\


\section{Improvements on the hardware for future deployment} \label{sec:improvements}
\subsection{TAXI v3.2}

The spectrum from the prototype station exhibits higher noise than the galactic background in the region between 50\,MHz to 100\,MHz (fig.~\ref{fig:spectrum}). The source of this noise was investigated and found to be generated by TAXI v3.0 itself.
Thus, the new central DAQ, TAXI v3.2, has a reduced RFI emission mainly due to a better shielding of the DC-DC converters. 
Other minor improvements, such as removing some of the unused components as well as increasing the bias voltage from 4.0\,V to 4.5\,V for the radioTads, were also included.
Another feature integrated in TAXI v3.2 is the possibility to turn on and off remotely the power to the antennas which allows better investigations on the internal noise of the DAQ system. 

\subsection{RadioTad v2}

It was noticed that the differences in gain between the channels of a single board were fluctuating above the \SI{10}{\percent} requirement for accurate energy and $X_\text{max}$ reconstructions of air showers. Furthermore, a phase difference between the channels was also observed. Although less crucial for cosmic-ray physics, it needed to be improved in order obtain a more uniform traces. To solve those issues, the following changes in the radioTad design were made :
\begin{itemize}
    \setlength\itemsep{-0.5em}
    \item A splitter \textit{SCA-4-10+} was added to fan-out the signal while keeping the impedance constant.
    \item The use of a new bias-tee \textit{TCBT-2R5G+} helped to reduce the size of the components on the board which led to an easier routing and a cost reduction.
    \item The layout of the board was completely redesigned with a special care for signal routing (length and impedance).
\end{itemize}

\begin{figure}
\centering
\begin{subfigure}{.5\textwidth}
  \centering
  \includegraphics[width=0.95\linewidth]{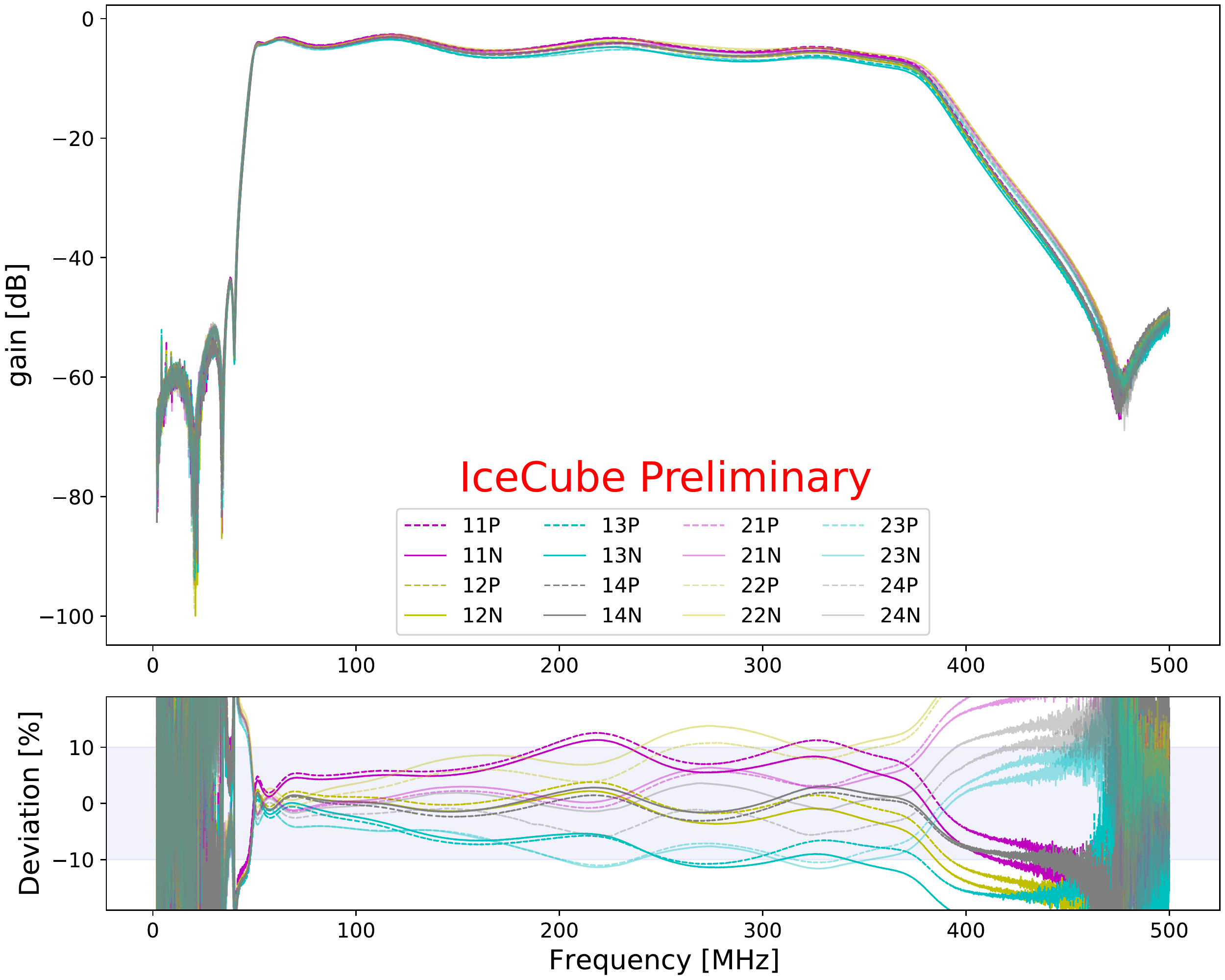}
  \caption{}
  \label{fig:radtadv1}
\end{subfigure}%
\begin{subfigure}{.5\textwidth}
  \centering
  \includegraphics[width=0.95\linewidth]{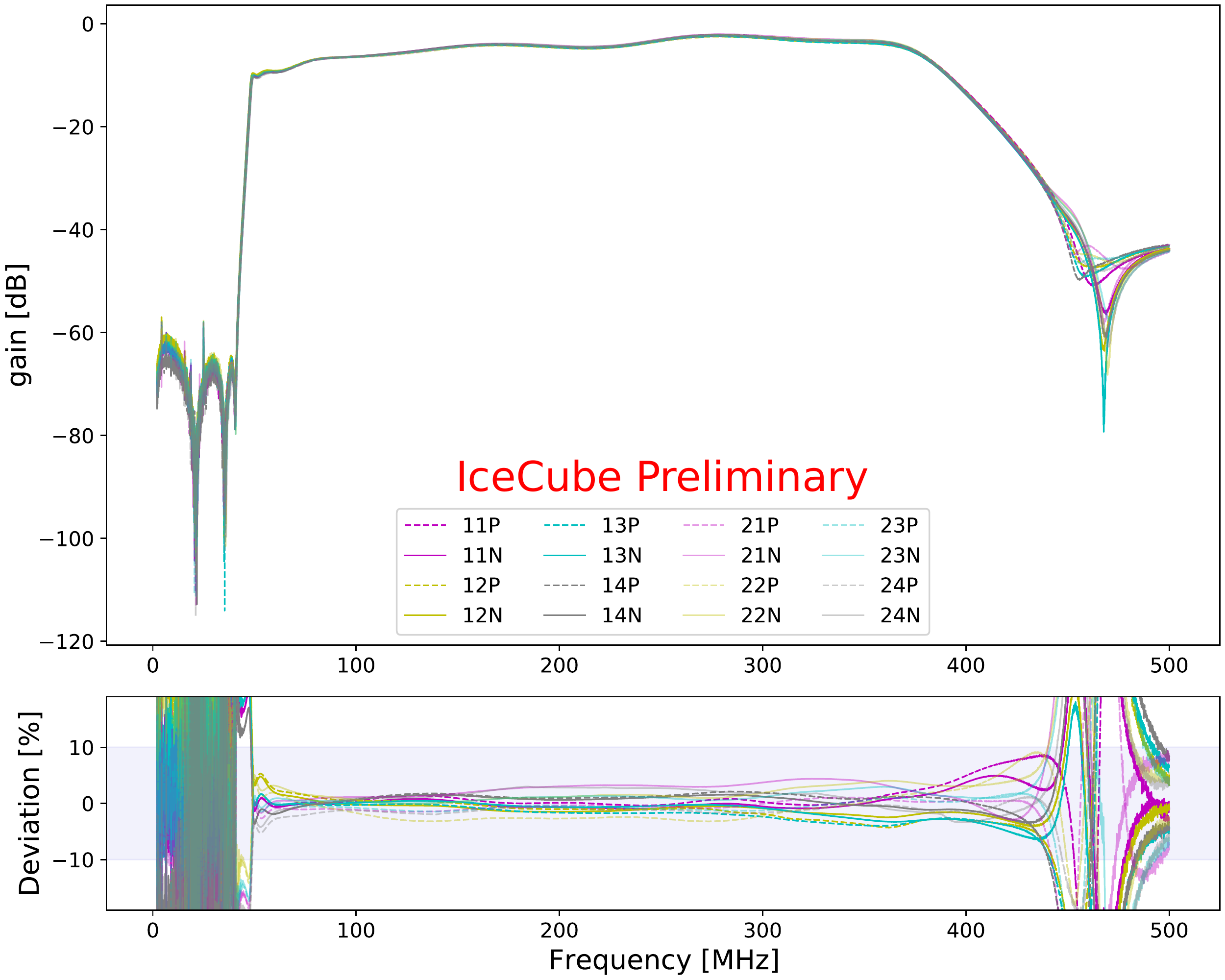}
  \caption{}
  \label{fig:radtadv2}
\end{subfigure}
\caption{a) Gain of the radioTad v1. b) Gain of the radioTad v2. The 16 lines represent each signal of the differential pairs.
The naming N and P in the legend refers to the two signals in the differential pairs and the first number to the antenna's polarisation. The effective total gain at the output of the TAXI, when the pairs are added together is \SI{6}{\decibel} higher. 
}
\label{fig:radtadGain}
\end{figure}

Fig.~\ref{fig:radtadGain} shows the reduced deviation of the gains between the 16 different channels. Each differential pair is taken individually for this measurement in order to closely observe their behavior. One can observe that the deviation of the old board (fig.~\ref{fig:radtadv1}) is above the \SI{10}{\percent} band whereas for the new board (fig.~\ref{fig:radtadv2}) it is located well below the \SI{10}{\percent} band.
Furthermore, the gain curve is flatter. In this figure, one can also see the sharp cut-off of the high-pass filter and the more relaxed cut-off of the low-pass filter.

\begin{wrapfigure}[11]{r}{0.50\textwidth}
\vspace{-10pt}
\centering
\includegraphics[width=\linewidth]{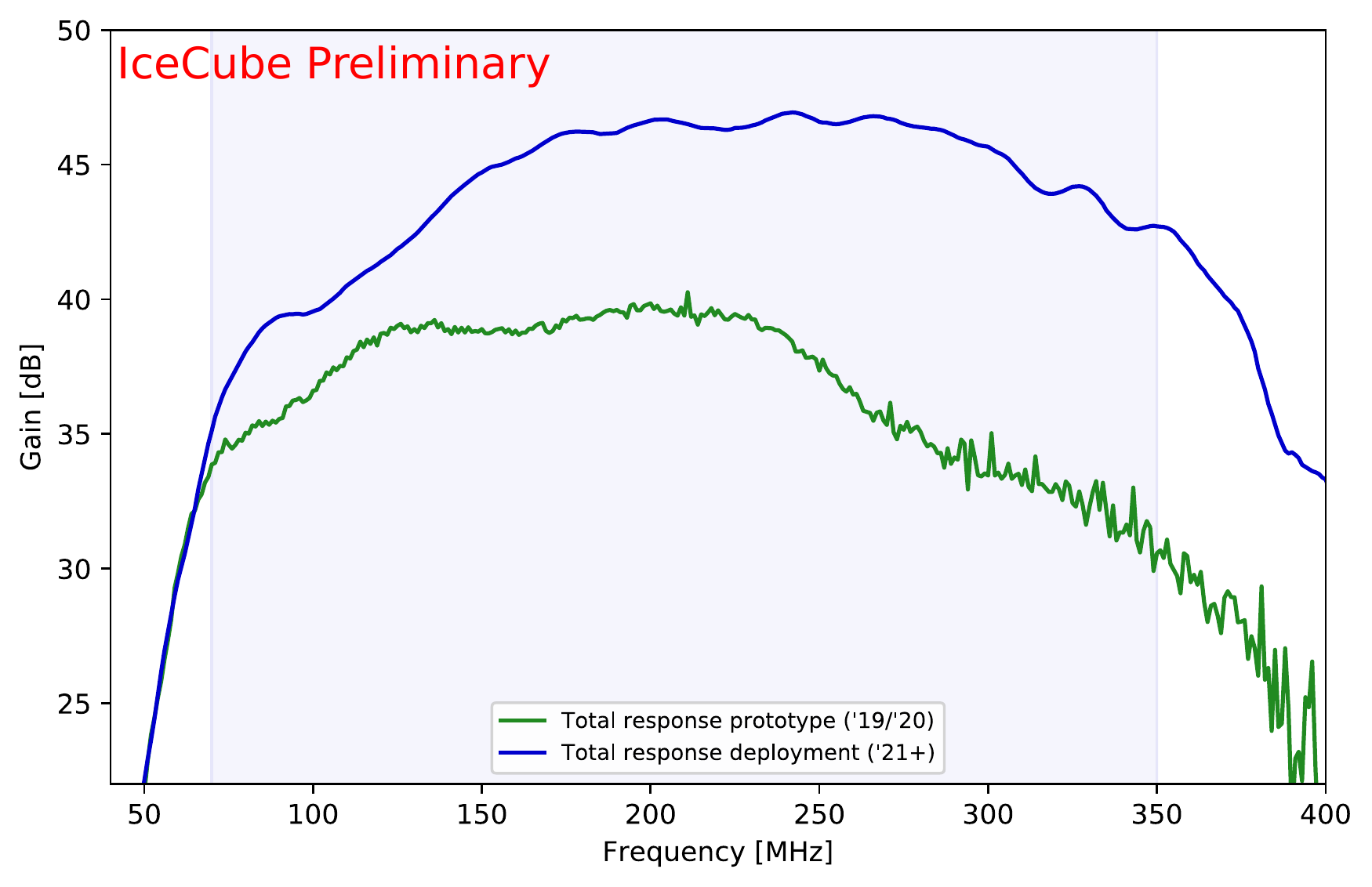}
\caption{Total gain comparison of the prototype station in green versus deployment station in blue.}
\label{fig:gainComparison}
\end{wrapfigure}
Fig.~\ref{fig:gainComparison} shows the gain of the prototype station (TAXI v3.0 + radioTad v1) compared to the improved electronics (TAXI v3.2 + radioTad v2).
One can see that especially at higher frequencies the amplification is improved and the gain is more consistent across the entire frequency band of interest. 
One can also notice the lower than average amplification in the lower frequencies (around \SI{100}{\mega\hertz}). 
Although TAXI v3.2 is better shielded, residual RFI will be attenuated by this lower gain.

\section{Conclusion and outlook}
The IceCube Surface Array Enhancement is designed to mitigate the effect of snow accumulation over the IceTop Cherenkov tanks, to improve the cosmic-ray measurements and to serve as R\&D for IceCube-Gen2.
The deployed prototype station has demonstrated that these goals can be achieved and proves the current design.
Further improvements of the hardware will enable the array to work optimally in the harsh South Pole environment.

\section{Acknowledgements}
We acknowledge the support by the Doctoral School "Karlsruhe School of Elementary and Astroparticle Physics: Science and Technology",  Dr. E. de Lera Acedo from the SKA Collaboration in addition to the IceCube acknowledgment list. Also, this project has received funding from the European Research Council (ERC) under the European Union's Horizon 2020 research and innovation programme (grant agreement No 802729).

\bibliographystyle{ICRC}
\bibliography{references}

\clearpage
\section*{Full Author List: IceCube Collaboration}




\scriptsize
\noindent
R. Abbasi$^{17}$,
M. Ackermann$^{59}$,
J. Adams$^{18}$,
J. A. Aguilar$^{12}$,
M. Ahlers$^{22}$,
M. Ahrens$^{50}$,
C. Alispach$^{28}$,
A. A. Alves Jr.$^{31}$,
N. M. Amin$^{42}$,
R. An$^{14}$
% ...
\\

\noindent
% ...
$^{12}$ Universit{\'e} Libre de Bruxelles, Science Faculty CP230, B-1050 Brussels, Belgium \\
% ...
$^{14}$ Department of Physics and Laboratory for Particle Physics and Cosmology, Harvard University, Cambridge, MA 02138, USA \\
% ...
$^{17}$ Department of Physics, Loyola University Chicago, Chicago, IL 60660, USA \\
$^{18}$ Dept. of Physics and Astronomy, University of Canterbury, Private Bag 4800, Christchurch, New Zealand \\
% ...
$^{22}$ Niels Bohr Institute, University of Copenhagen, DK-2100 Copenhagen, Denmark \\
% ...
$^{28}$ D{\'e}partement de physique nucl{\'e}aire et corpusculaire, Universit{\'e} de Gen{\`e}ve, CH-1211 Gen{\`e}ve, Switzerland \\
% ...
$^{31}$ Karlsruhe Institute of Technology, Institute for Astroparticle Physics, D-76021 Karlsruhe, Germany  \\
% ...
$^{42}$ Bartol Research Institute and Dept. of Physics and Astronomy, University of Delaware, Newark, DE 19716, USA \\
% ...
$^{50}$ Oskar Klein Centre and Dept. of Physics, Stockholm University, SE-10691 Stockholm, Sweden \\
% ...
$^{59}$ DESY, D-15738 Zeuthen, Germany \\
% ...

\subsection*{Acknowledgements}

\noindent
USA {\textendash} U.S. National Science Foundation-Office of Polar Programs,
U.S. National Science Foundation-Physics Division,
U.S. National Science Foundation-EPSCoR,
Wisconsin Alumni Research Foundation,
Center for High Throughput Computing (CHTC) at the University of Wisconsin{\textendash}Madison,
Open Science Grid (OSG),
Extreme Science and Engineering Discovery Environment (XSEDE),
Frontera computing project at the Texas Advanced Computing Center,
U.S. Department of Energy-National Energy Research Scientific Computing Center,
Particle astrophysics research computing center at the University of Maryland,
Institute for Cyber-Enabled Research at Michigan State University,
and Astroparticle physics computational facility at Marquette University;
Belgium {\textendash} Funds for Scientific Research (FRS-FNRS and FWO),
FWO Odysseus and Big Science programmes,
and Belgian Federal Science Policy Office (Belspo);
Germany {\textendash} Bundesministerium f{\"u}r Bildung und Forschung (BMBF),
Deutsche Forschungsgemeinschaft (DFG),
Helmholtz Alliance for Astroparticle Physics (HAP),
Initiative and Networking Fund of the Helmholtz Association,
Deutsches Elektronen Synchrotron (DESY),
and High Performance Computing cluster of the RWTH Aachen;
Sweden {\textendash} Swedish Research Council,
Swedish Polar Research Secretariat,
Swedish National Infrastructure for Computing (SNIC),
and Knut and Alice Wallenberg Foundation;
Australia {\textendash} Australian Research Council;
Canada {\textendash} Natural Sciences and Engineering Research Council of Canada,
Calcul Qu{\'e}bec, Compute Ontario, Canada Foundation for Innovation, WestGrid, and Compute Canada;
Denmark {\textendash} Villum Fonden and Carlsberg Foundation;
New Zealand {\textendash} Marsden Fund;
Japan {\textendash} Japan Society for Promotion of Science (JSPS)
and Institute for Global Prominent Research (IGPR) of Chiba University;
Korea {\textendash} National Research Foundation of Korea (NRF);
Switzerland {\textendash} Swiss National Science Foundation (SNSF);
United Kingdom {\textendash} Department of Physics, University of Oxford.

\end{document}